\documentclass[twocolumn,showpacs,preprintnumbers,amsmath,amssymb]{revtex4}
\usepackage{mathrsfs}
\usepackage{bbm}

\usepackage{graphicx}
\usepackage{dcolumn}
\usepackage{bm}

\begin{document}

\title{Band-gap bowing and \emph{p}-type doping of (Zn, Mg, Be)O wide-gap
semiconductor alloys: a first-principles study}
\author{Hongliang Shi$^{1,}$}
\email{hlshi@semi.ac.cn}
\author{Yifeng Duan$^{1,2}$}
\affiliation{$^{1}$State Key Laboratory for Superlattices and
Microstructures, Institute of Semiconductors, Chinese Academy of
Sciences, P.O. Box 912, Beijing 100083, People's Republic of China
\\$^{2}$Department of Physics, School of Sciences, China University
of Mining and Technology, Xuzhou 221008, People's Republic of China
}%

\date{\today}
\begin{abstract}

Using a first-principles band-structure method and a special
quasirandom structure (SQS) approach, we systematically calculate
the band gap bowing parameters and \emph{p}-type doping properties
of (Zn, Mg, Be)O related random ternary and quaternary alloys. We
show that the bowing parameters for ZnBeO and MgBeO alloys are large
and dependent on composition. This is due to the size difference and
chemical mismatch between Be and Zn(Mg) atoms. We also demonstrate
that adding a small amount of Be into MgO reduces the band gap
indicating that the bowing parameter is larger than the band-gap
difference. We select an ideal N atom with lower \emph{p} atomic
energy level as dopant to perform \emph{p}-type doping of ZnBeO and
ZnMgBeO alloys. For N doped in ZnBeO alloy, we show that the
acceptor transition energies become shallower as the number of the
nearest neighbor Be atoms increases. This is thought to be because
of the reduction of \emph{p}-\emph{d} repulsion. The N$_{\rm{O}}$
acceptor transition energies are deep in the ZnMgBeO quaternary
alloy lattice-matched to GaN substrate due to the lower valence band
maximum. These decrease slightly as there are more nearest neighbor
Mg atoms surrounding the N dopant. The important natural valence
band alignment between ZnO, MgO, BeO, ZnBeO, and ZnMgBeO quaternary
alloy is also investigated.

\end{abstract}

\maketitle

\section{INTRODUCTION}
Wide band-gap II-VI semiconductors ZnO, MgO, and BeO have been
extensively investigated because of their great potential
applications in the spintronic and optoelectronic
devices\cite{r1,r2,r3}. Recently ZnBeO alloy has been proposed as
another wide-band gap oxide, and its band gap can be efficiently
engineered to vary from the ZnO band gap (3.4 eV) to that of BeO
(10.6 eV)\cite{r4}. Furthermore, ZnO based UV light emitting diodes
(LEDs) with an active layer region composed of ZnO/ZnBeO quantum
wells also have been fabricated\cite{r5,r6}. Practically alloying is
a good approach to perform band-gap engineering to extend the
available band gap, and the band gap \emph{E}$_{g}$ of the alloy
\emph{A$_{x}$B$_{1-x}$C} can be described as
\begin{eqnarray}
E_{g}(x)=xE_{g}(AC) + (1-x)E_{g}(BC)-bx(1-x),
\end{eqnarray}
where \emph{b} is called the band-gap (optical) bowing parameter,
and $E_{g}(AC)$ and $E_{g}(BC)$ are the band gap of the binary
constituents \emph{AC} and \emph{BC}, respectively. For most
semiconductor alloys, the bowing parameter \emph{b} is almost
independent of composition \emph{x}\cite{r7}. However, the bowing
parameters \emph{b} of Ga$_{1-x}$N$_{x}$As and Zn$_{1-x}$Te$_{x}$O
alloys are strongly composition dependent, indicating that theses
alloys contain some elements with large differences between the size
and atomic orbital energies. These alloys have attracted a great
deal of attention\cite{r8,r9}. It is of great interest therefore to
investigate how the bowing parameters \emph{b} and the band gaps of
ZnBeO, MgBeO, and ZnMgO to vary since the large differences between
Zn (Mg) and Be atoms also exist in these alloys.

It is well known that the \emph{p}-type doping bottleneck poses a
challenge for the full utilization of the wide band-gap oxides
mentioned above. Due to the low valence band maximum (VBM), the
transition energies of acceptor defects are always deep and the
acceptors are hard to ionize at normal operating temperature. Great
efforts have been made to overcome the \emph{p}-type difficulty, for
example, through the codoping approach, complex defect, and impurity
band\cite{r10,r11,r12}. Li \emph{et al.} also proposed that
replacing Zn with isovalent Mg or Be surrounding with N acceptor
defects may act to lower the acceptor transition energy through
reducing the kinetic \emph{p-d} repulsion\cite{r11}. In the ZnBeO
alloy, it is easy to form N$_{\rm{O}}$+nBe$_{\rm{Zn}}$ complexes;
thus the transition energy is expected to be shallower when the
N$_{\rm{O}}$ defect is surrounded by more Be neighboring atoms.
Therefore it is also interesting to systematically study the
transition energy of the N$_{\rm{O}}$ defect in the ZnBeO alloy.

In this work we systematically investigate ZnBeO, ZnMgO, and MgBeO
alloys, in order to understand their band structure properties. We
hope to show that the bowing parameters are large and dependent on
composition for Mg$_{1-x}$Be$_{x}$O and Zn$_{1-x}$Be$_{x}$O alloys,
and much larger even in the Be-rich region. However, the bowing
parameter \emph{b} is small for the Zn$_{1-x}$Mg$_{x}$O alloy. We
also suggest that introducing a small amount of Be into MgO will
reduce the band gap. We note that these phenomena can be explained
by the size and atomic orbital energies differences between the
constituents and isovalent defect levels in these systems.
Furthermore, we also show that N is a good \emph{p}-type dopant in
the Zn$_{0.75}$Be$_{0.25}$O alloy which may be a good \emph{p}-type
doping material. However, the transition energies of N$_{\rm{O}}$ in
the Zn$_{0.6875}$Mg$_{0.25}$Be$_{0.0625}$O quaternary alloy
lattice-matched to GaN are deep due to the lower VBM according to
our band offsets calculations.

\section{CALCULATIONAL DETAILS AND METHODS}
Our first-principles band-structure and total energies calculations
are performed using the density functional theory (DFT) within the
local-density approximation\cite{r13} (LDA) as implemented in the
Vienna ab initio simulation package (vasp) code\cite{r14}. The
electron and core interactions are included using the frozen-core
projected augmented wave approach\cite{r15}. The Zn 3\emph{d}
electrons are explicitly treated as valence electrons. The whole
electron wave function is expanded in plane waves up to a cutoff
energy of 400 eV. All the geometries are optimised by minimising the
quantum mechanical forces acting on the atoms. For the Brillouin
zone integration, we use the \emph{k} points that are equivalent to
the 4$\times$4$\times$4 Monkhorst-Pack special \emph{k}-point
meshes\cite{r150} in the zinc-blende (ZB) Brillouin zone.

In order to calculate the band structure parameters of
\emph{A$_{x}$B$_{1-x}$C} alloys, we adopted the more efficient
special quasirandom structure (SQS) approach\cite{r16,r17}. In this
approach, we use a smaller unit cell to model the random alloys,
where mixed-atom sites are occupied based on the physically most
relevant structure correction functions which are closest to the
exact values of an random alloy. In our calculation, we use the
2\emph{a}$\times$2\emph{a}$\times$2\emph{a} 64 atoms SQS at
\emph{x}= 0.0625, 0.25, 0.50, 0.75 and 0.9375 for the
Mg$_{1-x}$Be$_{x}$O, Zn$_{1-x}$Be$_{x}$O and Zn$_{1-x}$Mg$_{x}$O
ternary alloys.

\section{RESULTS}
\subsection{Related binary, ternary and quaternary alloys}
In order to investigate the electronic and structure properties of
the random alloys, it is essential to calculate the lattice
constants and band gap parameters of related ZnO, MgO, and BeO
binary alloys. Wei \emph{et al.} concluded that the band structures
of the ZB and WZ phases for ZnO (GaN) are very similar near the band
edge at $\Gamma$ point through systematically investigating the
relationships between the band gaps of ZB and WZ phase of
semiconductors\cite{r18}. All our calculations are performed for
cubic zinc-blende(ZB) alloys and our results are also applicable for
the ground-state WZ alloys similar to other' studies\cite{r19}.

Table I shows our calculated equilibrium lattice constant \emph{a},
bulk modulus \emph{B}, and band gaps at $\Gamma$ point for
zinc-blende structure ZnO, MgO, BeO. We also list the corresponding
experimental values of bulk modulus \emph{B} and band gaps (in
brackets) for their ground-state structure. Our calculated lattice
parameters for $\emph{a}_{\rm{ZnO}}$, $\emph{a}_{\rm{MgO}}$, and
$\emph{a}_{\rm{BeO}}$ are 4.506, 4.517, and 3.766 \AA, respectively,
which are in good agreement with the experimental and theoretical
values 4.47, 4.524, and 3.768 \AA.

\begin{table}
\caption{Calculated lattice constant \emph{a}, bulk modulus
\emph{B}, and band gap \emph{E$_{g}$} at $\Gamma$ for ZB binary
alloys ZnO, MgO, and BeO. The experimental values of \emph{B}, and
\emph{E$_{g}$} suggested in Ref. 7 are also listed in parentheses.}
\label{tab.1}
\begin{ruledtabular}
\begin{tabular}{lccr}
& ZnO & MgO & BeO \\
\hline
\emph{a}($\rm{\AA}$) & 4.506(4.47$^a$) & 4.517(4.524$^b$) & 3.766(3.768$^c$) \\
\emph{B}(Mbar) & 1.562(1.837)  &  1.554(1.603) & 2.354(2.244) \\
& & & 2.29$^c$\\
\emph{E$_{g}$}(eV)  & 0.70(3.4)  & 3.731(7.67)  &  7.852(10.585) \\
\end{tabular}
$^a$experimental value in Ref. 20\\
$^b$theoretical value in Ref. 21
$^c$theoretical value in Ref. 22\\
\end{ruledtabular}
\end{table}

Conventionally, the lattice constant of Zn$_{1-x}$Be$_{x}$O alloy is
described by the Vegard law\cite{r23}, that is
\begin{eqnarray}
a(x)=(1-x)a_{\rm{ZnO}} + xa_{\rm{BeO}}.
\end{eqnarray}
Note that the lattice mismatch between ZnO (MgO) and BeO is 19.6\%
(19.9\%). Consequently, the ZnBeO and MgBeO alloys are highly
strained and the lattice constants may deviate from the Vegard law.
First, we calculate the lattice constant by the Vegard law, then fit
the Murnaghan equation of state\cite{r24} at various lattice
constants around it. Figure 1 shows the lattice constant as a
function of concentration \emph{x} obtained from the Vegard law
(black dots) and fitted by the Murnaghan equation (red dots). We
note that (i) For the ZnBeO and MgBeO alloys, due to the large size
difference between the Zn (Mg) and Be atom, the lattice constants
calculated by LDA (by fitting the Murnaghan equation) are slightly
larger than those obtained from the Vegard law. This is because the
Zn-O (Mg-O) bond is hard to compress. Our calculated results also
show that nearly all the Be-O bonds are longer than those of Be-O in
the host BeO. The biggest deviation occurring at composition
\emph{x}=0.5 is 2.5\% for ZnBeO alloys. (ii) the Vegard law is a
good model for ZnMgO alloys due to the lattice match and the similar
size of Zn and Mg ions. In actuality, we use the lattice constants
obtained by fitting the Murnaghan equation in all our calculations
for ternary alloys.

\begin{figure}
\includegraphics[width=6.5cm]{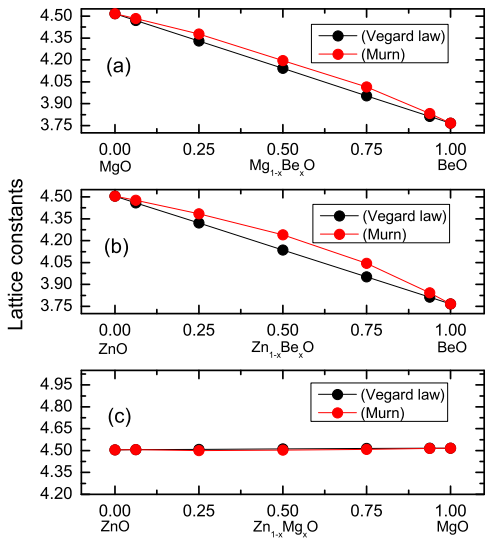} \caption{Comparison between
Vegard's law and LDA calculated (by fitting the Murnaghan equation
of state) lattice constants of Mg$_{1-x}$Be$_{x}$O,
Zn$_{1-x}$Be$_{x}$O and Zn$_{1-x}$Mg$_{x}$O alloys. The picture on
the left and right corresponds to two endpoint alloys (a) MgO
(\emph{x}=0) and BeO(\emph{x}=1), (b) ZnO (\emph{x}=0) and
BeO(\emph{x}=1), and (c) ZnO (\emph{x}=0) and MgO(\emph{x}=1),
respectively.} \label{fig.1}
\end{figure}

\subsection{The bowing parameters b of ZnMgO, ZnBeO, and MgBeO alloys}

We calculate the LDA band gaps to be 0.70, 3.731, 7.852 eV at
$\Gamma$ point for zinc-blende ZnO, MgO and BeO, respectively. Due
to the LDA underestimation of band gap, our values are smaller than
the experimental values\cite{r7} of 3.4, 7.67, 10.585 eV for their
ground-state phase, respectively. However, the error of bowing
parameter \emph{b} is small in Eq. 1 because the systematic band gap
error cancels out in deriving the bowing parameters through
comparing chemically identical systems in two different forms (AX
and BX)\cite{r25}. This has been confirmed by numerous
studies\cite{r9,r25}.

Table II presents the calculated band-gap bowing coefficients for
Mg$_{1-x}$Be$_{x}$O, Zn$_{1-x}$Be$_{x}$O and Zn$_{1-x}$Mg$_{x}$O
ternary alloys as a function of the composition \emph{x}. our
results show some chemical trends.

\begin{table}
\caption{Calculated band-gap bowing coefficient \emph{b} (eV) for
ternary alloys at different compositions.} \label{tab.2}
\begin{ruledtabular}
\begin{tabular}{lccccr}
&&\multicolumn{3}{c}{composition \emph{x}}\\\cline{2-6}
Ternary Alloys  & 0.0625 & 0.25 & 0.50 & 0.75 &0.9375 \\
Zn$_{1-x}$Mg$_{x}$O  & 0.8794 & 0.9797 & 1.4172 & 2.2827  & 4.3430 \\
Zn$_{1-x}$Be$_{x}$O  & 4.8215 & 5.8189 & 8.2116 & 12.8925 & 23.9648\\
Mg$_{1-x}$Be$_{x}$O  & 4.4951 & 5.2883 & 5.4772 & 7.2861
&18.0994 \\
\end{tabular}
\end{ruledtabular}
\end{table}

(i) The bowing parameters \emph{b} are large for ZnBeO and MgBeO
alloys. This is thought to be because of the atomic size difference
and the large chemical mismatch between Zn (Mg) and Be. However,
compared with ZnBeO and MgBeO the bowing parameters of ZnMgO alloys
are smaller as Zn and Mg are of similar size. The atomic size of Zn
is a little larger than that of Mg, thus the bowing parameters of
ZnBeO are also a little larger that those of MgBeO. From Be to Mg to
Zn, the atomic size difference increases as does the atomic
\emph{p}-orbital energy difference.

(ii) The bowing parameters are composition and large at Be-rich
region for ZnBeO and MgBeO alloys. Taking ZnBeO as an example, the
bowing parameter \emph{b} is 4.8215 eV in the Zn-rich region,
whereas in the Be-rich region the \emph{b} is 23.9648 eV. This is
because the size and chemical differences between Zn and Be are
large. Incorporation of a small amount of Zn into BeO at
substitutional sites creates a deep isovalent defect level inside
the band gap. Our direct calculations show that the position of the
isovalent defect level is at 0.66 eV below the conduction band
minimum (CBM). Furthermore the VBM is pushed up due to the
\emph{p-d} repulsion introduced by Zn 3\emph{d} orbital. This leads
to a large band-gap reduction and a large bowing parameter in the
Be-rich region. However, upon adding a small amount of Be into ZnO,
the VBM will shift downwards by a small amount due to the reduction
in \emph{p-d} repulsion. In this case the band gap increases a
little and the bowing parameter is small.

(iii) Using the bowing parameters and experimental band gaps, we
calculate the band gaps of ZnBeO, MgBeO, and MgBeO alloys at
different concentrations. The results are summarized in Figure 2. We
notice that although the band gap of BeO is larger than that of MgO,
incorporation of a small amount of Be into the MgO matrix will lead
to a reduction in the band gap, indicating that the bowing parameter
is larger than the band-gap difference. For Mg$_{0.75}$Be$_{0.25}$O,
the bowing parameter \emph{b} (5.2883 eV) is larger that the band
gap difference (2.915 eV).

\begin{figure}
\includegraphics[width=6.5cm]{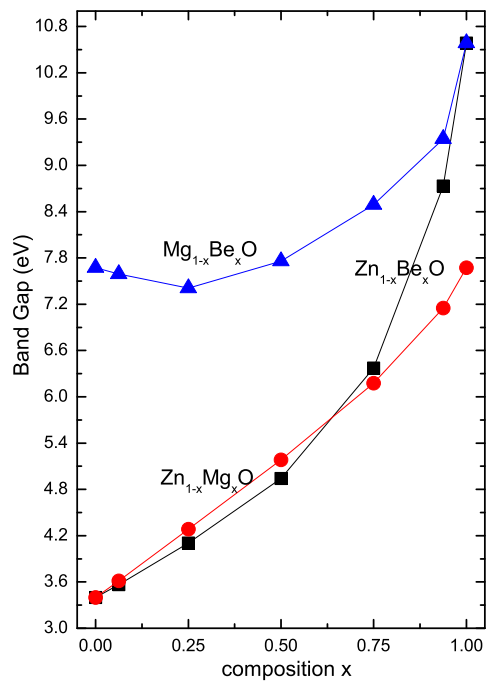} \caption{The predicted band gaps
of Mg$_{1-x}$Be$_{x}$O, Zn$_{1-x}$Be$_{x}$O and Zn$_{1-x}$Mg$_{x}$O
ternary alloys calculated at various compositions.} \label{fig.2}
\end{figure}

\subsection{The formation energies and transition energy levels
of N in ZnBeO and ZnMgBeO quaternary alloys}

\emph{P}-type doping of ZnO is difficult due to the low VBM
resulting from the largely electronegative characteristic of oxygen.
In order to lower the acceptor ionization energy extensive studies
have been performed. Park \emph{et al}.\cite{r10} showed that N is
the best \emph{p}-type dopant source among the group-I and V
elements. Li \emph{et al.}\cite{r11} also showed that replacing Zn
with Mg or Be surrounding the N$_{\rm{O}}$ substitutional defect can
decrease the acceptor transition level. In the ZnBeO ternary alloy,
it is easy to form N$_{\rm{O}}$+nBe$_{\rm{Zn}}$ complexes; thus we
expect the acceptor transition level becomes shallower when the
N$_{\rm{O}}$ is surrounded by more Be atoms. In the following study,
we choose the Zn$_{0.75}$Be$_{0.25}$O alloy approximately as the
experimental investigations do\cite{r5}, and we also choose the N as
dopant due to its low valence \emph{p}-orbital energy.

The ionization energy of an accepter (\emph{q}$<$0) with respect to
the VBM in the impurity limit is calculated by the following
procedure, further described in Refs. 26 and 27,
\begin{eqnarray}
\epsilon(0/q)=&&\{E(\alpha,q)-[E(\alpha,0)-q\epsilon^{k}_{D}(0)]\}/(-q)\nonumber\\
&&+[\epsilon^{\Gamma}_{D}(0)-\epsilon^{\Gamma}_{\rm{VBM}}(\rm{host})],
\end{eqnarray}
where $E(\alpha,q)$ and $E(\alpha,0)$ are the total energies of the
supercell at charge state $q$ and neutral, respectively, for defect
$\alpha$, $\epsilon^{k}_{D}(0)$ and $\epsilon^{\Gamma}_{D}(0)$ are
the defect levels at the special $k$ points (averaged) and at the
$\Gamma$ point, respectively, and
$\epsilon^{\Gamma}_{\rm{VBM}}$(host) is the VBM energy of the host
supercell at the $\Gamma$ point and is aligned using core electron
levels away from the defect. The first term on the right-hand side
of Eq. (5) determines the U energy parameter (including both the
Coulomb contribution and atomic relaxation contribution) of the
charged defects calculated at the special \emph{k} points, which is
the extra cost of energy after moving (-\emph{q}) charge from the
VBM of the host to the neutral defect level. The second term gives
the single-electron defect level at the $\Gamma$ point. For charged
defects, a uniform background charge is added to retain the global
charge neutrality of the periodic unit-cell.

The defect formation energy of a neutral defect is defined as
\begin{eqnarray}
\Delta
H(\alpha,0)=E(\alpha,0)-E(\rm{host})+n_{\rm{O}}\mu_{\rm{O}}+n_{\alpha}\mu_{\alpha},
\end{eqnarray}
where E(host) is the total energy of the host supercell without
defect $\alpha$; $\mu_{\rm{O}}$ and $\mu_\alpha$ are the chemical
potentials of constituents O and \emph{$\alpha$} (N) relative to the
element gas energy. The $n_{\rm{O}}$ ($n_{\rm{O}}$$>$0) and
$n_\alpha$ ($n_\alpha$$<$0) are the numbers of O and extrinsic
defects $\alpha$ (N). We calculate the chemical potentials for
O($\mu_{\rm{O}}$) and N($\mu_{\rm{N}}$), relative to gaseous O$_{2}$
and N$_{2}$.

Table III shows our calculated N$_{\rm{O}}$ acceptor formation
energies and $\epsilon$(0/-) transition energy levels for the ZnO
and Zn$_{0.75}$Be$_{0.25}$O alloy. For N in Zn$_{0.75}$Be$_{0.25}$O
alloy, we calculate the N$_{\rm{O}}$ at different sites. The
\emph{n}NN in Table III denotes that there are \emph{n} Be atoms in
the tetrahedral nearest neighbor (NN) sites centered around the N
atom. Our calculated results show the following  typical behaviours.

\begin{table}
\caption{Calculated formation energy $\Delta$H$_{f}$($\alpha$,0)
(eV) and transition energy level $\epsilon$(0/-) (eV) for
Zn$_{0.75}$Be$_{0.25}$O:N and ZnO:N are listed.} \label{tab.3}
\begin{ruledtabular}
\begin{tabular}{lcr}
 & $\Delta$H$_{f}$($\alpha$,0)& $\epsilon$(0/-) \\
 \hline
 0NN & 4.793 &  0.533\\
 1NN & 4.818 &  0.347\\
 2NN & 5.134 &  0.233\\
 ZnO & 4.661 &  0.362\\
\end{tabular}
\end{ruledtabular}
\end{table}

(i) The formation energy of neutral N$_{\rm{O}}$ defect in the
Zn$_{0.75}$Be$_{0.25}$O alloy increases when there are larger
numbers of Be atoms in the defect neighboring sites. This is because
the N$_{\rm{O}}$ substitution introduces a larger strain into the
Be-rich region than into the Zn-rich region. The formation energy of
N$_{\rm{O}}$ in the ZnO is smaller than that in the alloy. This is
because the volume of ZnO is about 0.9\% greater than that of the
Zn$_{0.75}$Be$_{0.25}$O alloy, indicating that an N$_{\rm{O}}$
defect in pure ZnO introduces a smaller strain.

(ii) For N$_{\rm{O}}$ in ZnO, our calculated N$_{\rm{O}}$
$\epsilon$(0/-) acceptor level is at 0.362 eV above the VBM, which
agrees well with previous first-principles
calculations\cite{r10,r11}. For the case of the
Zn$_{0.75}$Be$_{0.25}$O alloy, we note that the position of
N$_{\rm{O}}$ $\epsilon$(0/-) acceptor level is dependent on the site
of the defect. It decreases as the number of N nearest neighbouring
atoms increases, from 0.533 eV when N has no nearest neighboring
(0NN) Be atoms to 0.233 eV when there are 2 NN Be atoms. This is
thought to be because the acceptor level is lowered by the reduction
in the \emph{p}-\emph{d} repulsion. A more detailed explanation is
presented. For ZnO, the VBM state consists mostly of
\emph{p}$_{\rm{O}}$, \emph{p}$_{\rm{Zn}}$, and \emph{d}$_{\rm{Zn}}$
orbitals. For an N$_{\rm{O}}$ acceptor in ZnO, the defect level
mainly consists of N impurity valence \emph{p} orbitals, and the
acceptor level is pushed up by the \emph{p-d}
(\emph{p}$_{\rm{N}}$-\emph{d}$_{\rm{Zn}}$) repulsion introduced by
the same symmetry. However, for Be, it has no occupied 2\emph{p} and
3\emph{d} orbitals. If Zn is substituted by Be, the \emph{p-d}
(\emph{p}$_{\rm{N}}$-\emph{d}$_{\rm{Zn}}$) repulsion will decreases
and the N$_{\rm{O}}$ acceptor transition level becomes shallow. We
also note that the \emph{p-p} repulsion can push the acceptor level
further down, although it is weaker compared with \emph{p-d}
repulsion because the \emph{p-p} (\emph{p-d}) repulsion is inversely
proportional to the energy difference between the
\emph{p}$_{\rm{N}}$ and \emph{p}$_{\rm{Be}}$ (\emph{p}$_{\rm{N}}$
and \emph{d}$_{\rm{Zn}}$) valence orbitals\cite{r280}. Similarly, in
the Zn$_{0.75}$Be$_{0.25}$O alloy, from 0NN to 2NN, the
\emph{p}$_{\rm{N}}$-\emph{d}$_{\rm{Zn}}$ repulsion decreases and the
\emph{p-p} (\emph{p}$_{\rm{N}}$-\emph{p}$_{\rm{Be}}$) repulsion
increases; thus, the N$_{\rm{O}}$ acceptor level becomes
increasingly shallower. In 0NN, the N$_{\rm{O}}$ acceptor level in
the Zn$_{0.75}$Be$_{0.25}$O alloy is deeper than that in ZnO due to
the lower VBM (see below).

(iii) According to Eq. 4, we see that the chemical potentials have
large effect on the defect formation energy. Practically, the
chemical potentials can vary in a range because of some
thermodynamic limits\cite{r29}. The high defect formation energy
leads to a low dopant solubility, which may be overcome by
non-equilibrium growth methods\cite{r29}. If the formation energy of
N$_{\rm{O}}$ in Zn$_{0.75}$Be$_{0.25}$O alloy is reduced so that it
is smaller than that in ZnO by controlling the chemical potentials
of N and O during the growth process, there are more N atoms
residing in the 1NN and 2NN sites with low acceptor transition
levels. Based on this, we predict that ZnBeO alloy may be a better
\emph{p}-type doping material than ZnO.

We have also investigated the electronic structures and
\emph{p}-type doping properties of the ZnMgBeO quaternary alloy
lattice-matched to GaN substrate. For quaternary alloys, the band
gap and lattice constant can be individually tuned. For
Zn$_{1-x-y}$Mg$_{y}$Be$_{x}$O alloy, the lattice constant can be
described by
\begin{eqnarray}
a(x,y)=(1-x-y)a_{\rm{ZnO}} + ya_{\rm{MgO}} + xa_{\rm{BeO}}.
\end{eqnarray}
For the Zn$_{22}$Mg$_{8}$Be$_{2}$O$_{32}$ alloy, we find the Vegard
law is well obeyed. We also note that the lattice constants of
Zn$_{22}$Mg$_{8}$Be$_{2}$O$_{32}$ quaternary alloy and zinc-blende
GaN are 4.461 and 4.462 \AA\cite{r30}, respectively, thus the
lattices match well.

The band gap of the Zn$_{1-x-y}$Mg$_{y}$Be$_{x}$O quaternary alloys
is a function of compositions \emph{x} and \emph{y} and can be
described by a second order expansion in \emph{x} and
\emph{y}\cite{r301},
\begin{eqnarray}
E_{g}(x,y)=&(1-x-y)E_{g}(\rm{ZnO})+\emph{y}\emph{E}_{g}(\rm{MgO})\nonumber\\
           &+\emph{x}\emph{E}_{g}(\rm{BeO})-\emph{b}_{\rm{ZnMgO}}y(1-y)\nonumber\\
           &-b_{\rm{ZnBeO}}x(1-x)-\emph{b}_{\emph{xy}}\emph{xy}.
\end{eqnarray}
In order to get \emph{b}$_{xy}$ approximately, we use the SQS at
(\emph{x},\emph{y})=(0.0,0.5),(0.5,0.0), and (0.5,0.5). The
parameter \emph{b}$_{xy}$ is obtained using
\emph{b}$_{xy}$=\emph{b}$_{\rm{ZnBeO}}$-\emph{b}$_{\rm{ZnMgO}}$-\emph{b}$_{\rm{MgBeO}}$
under a quadratic approximation. Our calculated \emph{b}$_{xy}$ is
1.317 eV. Using the experimental band gaps and Eq. 6, we predict
that the band gap of Zn$_{22}$Mg$_{8}$Be$_{2}$O$_{32}$ alloy is
4.149 eV.

Table IV shows calculated N$_{\rm{O}}$ acceptor formation energies
and $\epsilon$(0/-) transition energy levels for
Zn$_{22}$Mg$_{8}$Be$_{2}$O$_{32}$ alloy. The \emph{n}NN in Table IV
denotes that there are \emph{n} Mg atoms in the tetrahedral nearest
neighbor (NN) sites centered around the N atom. The 0NN1Be denotes
that N atom has one Be atom and without Mg atoms in the NN sites.
From 0NN to 3NN, the $\epsilon$(0/-) decreases slightly less than
that of the Zn$_{0.75}$Be$_{0.25}$O alloy. This is because the
\emph{p}$_{\rm{Mg}}$ orbital energy level is higher than that of
\emph{p}$_{\rm{Be}}$\cite{r280}, thus the \emph{p-p} repulsion
pushing down the acceptor level is weaker in the quaternary alloy.
The $\epsilon$(0/-) is smaller in 0NN1Be than that in 0NN, which is
also consistent with our findings for N$_{\rm{O}}$ in the
Zn$_{0.75}$Be$_{0.25}$O alloy. We note that the N$_{\rm{O}}$
$\epsilon$(0/-) transition energy level is deeper than that in ZnO
because of the large downward shift of the VBM according to our band
alignment calculation (see below). The CMB of
Zn$_{22}$Mg$_{8}$Be$_{2}$O$_{32}$ alloy is as low as that of ZnO,
indicating that it is also can be \emph{n}-type doped easily.
\begin{table}
\caption{Calculated formation energy $\Delta$H$_{f}$($\alpha$,0)
(eV) and transition energy level $\epsilon$(0/-) (eV) for ZnMgBeO:N
are listed.} \label{tab.4}
\begin{ruledtabular}
\begin{tabular}{lcr}
 & $\Delta$H$_{f}$($\alpha$,0)& $\epsilon$(0/-) \\
 \hline
 0NN & 4.773 & 0.546\\
 1NN & 5.176 & 0.489\\
 2NN & 5.611 & 0.458\\
 3NN & 6.033 & 0.446 \\
0NN1Be&4.978 & 0.477\\
\end{tabular}
\end{ruledtabular}
\end{table}

\section{The natural valence band alignment}
Band offset is an important parameter in the heterostructures. We
shall describe how we investigated the natural band alignment by
adopting the approach proposed by Wei\cite{r31}. This method is also
widely used in photoemission core level spectroscopy\cite{r32}.
According to Wei's method\cite{r31}, the band offset is given by
\begin{eqnarray}
\Delta E_{v}(AX/BY)= \Delta E_{v,C^{'}}^{BY}-\Delta
E_{v,C}^{AX}+\Delta E_{C,C^{'}}^{AX/BY}.
\end{eqnarray}
Here,
\begin{eqnarray}
\Delta E_{v,C}^{AX}=E_{v}^{AX}-E_{C}^{AX}
\end{eqnarray}
(and similarly for $\Delta E_{v,C^{'}}^{BY}$) are the energy
separations between core level (C) and valence band maximum energy
for pure AX (and similarly for pure BY), while
\begin{eqnarray}
\Delta E_{C,C^{'}}^{AX/BY}=E_{C^{'}}^{BY}-E_{C}^{AX}
\end{eqnarray}
is the difference in core level binding energy between AX and BY at
the AX/BY superlattice. We calculate the $\Delta E_{v,C}^{AX}$ and
$\Delta E_{v,C^{'}}^{BY}$ at the equilibrium zinc-blende lattice
constants for AX and BY, respectively. We use the average lattice
constant $\bar{a}$ (between AX and BY)  to calculate the $\Delta
E_{C,C^{'}}^{AX/BY}$ in the AX/BY superlattice. A more detailed
description of the procedure can be found in Ref. 32.

The natural conduction band minimum (CBM) alignment is calculated by
\begin{eqnarray}
\Delta E_{c}(AX/BY)=\Delta E_{v}(AX/BY)+\Delta E_{g}({AX/BY}),
\end{eqnarray}
where $\Delta E_{g}(AX/BY)$ is the experimental band gap difference
between the AX and BY alloys.

\begin{figure}
\includegraphics[width=6.5cm]{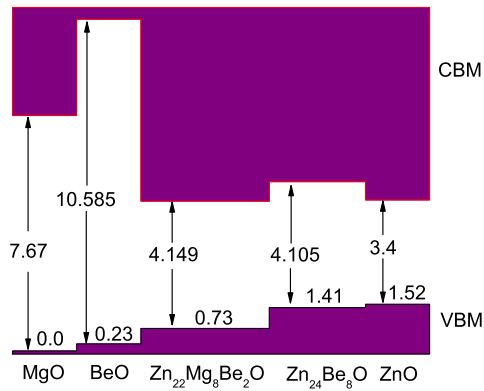} \caption{The calculated natural
band alignment of MgO, BeO, Zn$_{22}$Mg$_{8}$Be$_{2}$O$_{32}$,
Zn$_{24}$Be$_{8}$O$_{32}$ and ZnO alloys.} \label{fig.3}
\end{figure}

Our calculated natural band alignment is shown in Fig. 3. For the
VBM alignment, the VBM of ZnO is higher than that of MgO and BeO.
This is because the \emph{p-d} ($p_{\rm{O}}-d_{\rm{Zn}}$) repulsion
pushes up the VBM, whereas Mg and Be have no active \emph{d}
electrons. According to the common-anion rule\cite{r31}, the valence
band offset is small between common-anion systems. However, the VBM
of BeO is higher than that of MgO. This is because BeO shows strong
covalency, resulting in the kinetic-energy-induced valence band
broadening\cite{r33}. If we assume that the VBM varies linearly as a
function of the alloy composition, Zn$_{24}$Be$_{8}$O$_{32}$ would
have a VBM approximately 0.32 eV lower than that of ZnO. However,
our direct calculation shows that VBM of Zn$_{24}$Be$_{8}$O$_{32}$
(Zn$_{0.75}$Be$_{0.25}$O) alloy is only about 0.11 eV lower than
that of ZnO due to the large \emph{p-d} repulsion. In order to
investigate the natural conduction band alignment, we use the
experimental band gaps demonstrated in Fig. 3. We see that the
Zn$_{0.75}$Be$_{0.25}$O alloy acts as a good barrier material for
ZnO-based UV LEDs. The $\Delta E_{C}$ and $\Delta E_{V}$ have values
of approximately 0.60 and 0.11 eV, respectively, which are large
enough to strongly confine the electrons and holes well\cite{r34}.
The quaternary alloy Zn$_{22}$Mg$_{8}$Be$_{2}$O$_{32}$ has a CBM as
low as that of ZnO, indicating that it can be \emph{n}-type doped
easily. By contrast \emph{p}-type doping is more difficult due to
the low VBM which is also confirmed by our study of \emph{p}-type
doping above. Based on these chemical behaviours, we can tailor the
band offset to practically perform the band-engineering as desired.
We hope that our results will be helpful for the applications of
optoelectronic devices.

\section{SUMMARY}
We have investigated the band gap bowing parameters and
\emph{p}-type doping properties of (Zn, Mg, Be)O related random
ternary and quaternary alloys using first-principles band-structure
methods. The bowing parameters are large and dependent on
composition for ZnBeO and MgBeO alloys due to their large size and
chemical mismatch. We also show that the band gap of MgO will reduce
on adding a small amount of Be into MgO. The acceptor transition
energy level becomes shallower when N has more Be nearest neighbors,
thus ZnBeO (Zn$_{0.75}$Be$_{0.25}$O) may be a good \emph{p}-type
material. The electronic and \emph{p}-type doping properties of
ZnMgBeO quaternary alloy lattice-matched to GaN substrate are also
studied. According to our natural band alignment calculations,
\emph{n}-type doping of Zn$_{22}$Mg$_{8}$Be$_{2}$O$_{32}$ should be
easy due to the low CBM. We hope our conclusions have applications
to optoelectronic devices.

\acknowledgments This work was supported by the National Basic
Research Program of China (973 Program) grant No. G2009CB929300 and
the National Natural Science Foundation of China under Grant Nos.
60521001 and 60776061.


\begin{thebibliography}{0}
\bibitem{r280}
Our calculated (semi-relativistic) atomic LDA valence orbital
energies for $\epsilon_{p}^{\rm{N}}$-$\epsilon_{p}^{\rm{Be}}$,
$\epsilon_{p}^{\rm{N}}$-$\epsilon_{d}^{\rm{Zn}}$ and
$\epsilon_{p}^{\rm{Mg}}$-$\epsilon_{p}^{\rm{Be}}$ are -5.137, 3.181
and 0.728 eV, respectively.
\end{thebibliography}
\end{document}